\documentclass[prl,aps,twocolumn, floatfix,showpacs]{revtex4}
\usepackage{amsmath,bm,graphicx}
\usepackage{amssymb}
\usepackage{amsfonts}
\usepackage{graphicx}
\usepackage{hyperref}
\usepackage{latexsym}

\def\be{\begin{equation}}
\def\ee{\end{equation}}
\def\bea{\begin{eqnarray}}
\def\eea{\end{eqnarray}}

\begin{document}

\title{Jamming transition in a driven lattice gas}
\author{Sudipto Muhuri}
\affiliation {Department of Physics, University of Pune, Ganeshkhind, Pune 411007, India}


\begin{abstract} 
We study a two-lane driven lattice gas model with oppositely directed species of particles moving on two periodic lanes with {\it correlated} lane switching processes. While the overall density of individual species of particles is conserved in this system, the particles are allowed to switch lanes with finite probability only when oppositely directed species meet on the same lane. This system exhibits an unique behavior, wherein phase transition is observed between an homogeneous {\it absorbing} phase, characterized by complete segregation of oppositely directed particles between the two lanes, and a jammed phase. The transition is accompanied by a finite drop of current in the lattice, emergence of a cluster comprising of both species of particles in the jammed phase, and is determined by the interplay of the relative rates of translation of particles on the same lane and their lane switching rates. These findings may have interesting implications for understanding the phenomenon of jamming in microtubule filaments observed in context of axonal transport.
\end{abstract}

\maketitle
Unlike one-dimensional(1D) systems in thermal equilibrium, 1D and quasi 1D driven diffusive systems have a stationary state behavior, characterized  by macroscopic currents. These systems can exhibit spontaneous symmetry breaking, phase separation and {\it condensation}, resulting in very rich and complex phase diagrams which is in contrast to 1D equilibrium lattice gas models \cite{schutzrev, bridge2,evansrev, sriram,evans,arndt, popkov, kafri1,kafri2, satya}. One of the motivation for studying such systems is their amenability in providing a framework for studying varied class of driven biological processes, such as transport on biofilaments \cite{menon,freylet,ignapre}, growth of fungal filaments \cite{fungievans, fungiepl}, transport across biomembranes \cite{choubio}, among others. 

In this letter, we study a periodic two lane driven lattice gas system with oppositely directed species and {\it conserved} overall density of individual species \cite{ignajstat}. This system  incorporates bidirectionality and {\it correlated} lane switching processes, wherein oppositely directed species can switch lanes with a finite probability only when they encounter each other and not otherwise. Such {\it correlated} lane switching mechanism fundamentally alters the steady state properties. We find that the system exhibits an unique behavior, wherein a phase transition is observed between a jammed {\it clustered} phase to an homogeneous {\it absorbing} phase, characterized by complete segregation of oppositely directed particles between the two lanes. The jammed phase in each lane is characterized by a large cluster comprising of both species of particles and no vacancies, which is surrounded by a region of single species fluid phase in rest of the lane. This phase transition is  distinct from  phase transitions observed for other multi-species driven lattice gas models with conserved particle densities, such as the ones discussed in \cite{arndt}, where a transition from an homogeneous to a phase segregated state of the two species is observed, or where the transition from a two species homogeneous phase to a {\it condensate} phase is observed \cite{kafri2}. While transition from a jammed state to free flowing state of particles has been observed for driven systems which exchange particles with environment \cite{lipoepl,santen}, both the mechanism and the nature of the steady state is very different for our case, owing to the constraint of particle number conservation. Further, the behavior of this system is in contrast to other two lane models \cite{juhasz1, pronina1, frey2lane}, and to a closely related periodic two lane model with {\it conserved} particle number and {\it uncorrelated} lane switching mechanism \cite{korniss}, where the steady state is characterized by large but finite size clusters and no phase transition is observed in the thermodynamic limit of $N\rightarrow\infty$ \cite{korniss,kafri1}.

From a biological standpoint, motor protein driven bidirectional transport of cellular cargoes on multiple parallel filaments have been observed, for example in context of axonal transport in neurons \cite{welte,roop}. Filament switching of the motors between neighbouring microtubule (MT) filaments is also seen \cite{ross}. For neurons in brain cell, it has been suggested that neurodegenerative diseases like Alzheimer's, results from blockage and jamming of the transport machinery comprising of microtubules and motors \cite{jam1, jam2}. Thus it becomes imperative to understand the physical origin of jamming in such situations. Various alternative scenarios giving rise to jamming and impaired transport on microtubule filaments have been proposed based on experimental studies. Broadly they fall in two different categories. The first category focuses upon the role of the microtubules themselves and it has been proposed that jamming occurs either due to polar reorientation of the microtubule filaments along axons \cite{shemesh} or due to excess microtubule polymerization and bundling, followed by the degeneration of the microtubules \cite{thies}. The other category identifies the role played by the molecular motors in causing jam. In particular,  some studies have suggested that the jams occur either due to high motor density and low dissociation rates at MT filament ends \cite{leduc} or due to changes in the motor processivity on the filament \cite{jam2}. In fact one of the strategies employed for curing neurodegenerative diseases focuses upon removal of the jam by altering the movement of motors on the filament \cite{jam2}. In this context, the minimal model discussed here mimics the interplay of motor movement and filament switching processes of motors and illustrates a plausible physical mechanism which can in principle give rise to a transition between a jammed state to a freely flowing state. 

The model that we study comprises of two periodic lattice of length $L$ with $N$ sites.  Each lattice site can either be empty or it can be occupied by a  $(+)$ particle or a $(-)$ particle. In each lattice,  $(+)$  particle hops to the right  with a rate $\alpha$ if the adjacent site to the right is vacant. Similarly,  a $(-)$ particle hops to the left  with  the same rate $\alpha$ if the adjacent site to the left is unoccupied.  For a $(+)$ particle on a lattice site $i$, if the neighbouring site to the right on the same lattice  is occupied by a $(-)$ particle, then two different  processes can take place;  either the $(+)$ particle hops to the neighbouring site at $i + 1$ on the same lattice while the neighbouring $(-)$ particle hops to the site $i$  with a rate $\beta$, or the $(+)$ particle in lattice 1(2) switches with a rate $\gamma_{12}(\gamma_{21})$ to the corresponding site $i$ on the other lattice if that site is vacant. Similarly for $(-)$ particles if the neighbouring site to the left is occupied with a $(+)$ particle, then the $(-)$ particle hops to the neighbouring site at $i - 1$ on the same lattice while the neighbouring $(+)$ hops to the site $i$ with rate $\beta$, or the $(-)$ particle switches to the other lattice with a rate $\mu_{12} ( \mu_{21})$ if the site $i$ of the opposite lane is vacant.

We restrict ourselves to studying the system for which the total conserved density of $(+)$ and $(-)$ are equal, so that $\rho_{+}  = \rho_{-}  = \rho_{o}$, where $\rho_{+}$ and $\rho_{-}$ are the conserved total density of $(+)$ and $(-)$ particles respectively. We choose $\gamma_{12}= \mu_{21} = \gamma$, $\gamma_{21} = \mu_{12} = \mu$ with $\mu > \gamma$ and set $\alpha =1$, expressing the other rates in terms of it. We study the system using a combination of Mean Field (MF) analysis and Monte Carlo (MC) simulations \footnote{In MC simulation, we wait for an initial transient $\geq 20000 \frac{N}{\alpha}$ swaps. Averaging is done typically over $10^{3}-10^{4}$ time swaps with a period  $\geq 20 \frac{N}{\alpha}$.}. 

To characterize the steady state, we analyze the density and current profile of the system in steady state. We denote the mean densities as a function of relative position $x$ along the lanes as $p_{1}(x)$, $p_{2}(x)$, $n_{1}(x)$ and $n_{2}(x)$, corresponding to density of $(+)$ in lane $1$, $(+)$ in lane $2$, $(-)$ in lane $1$, and $(-)$ in lane $2$ respectively. We also define an order parameter $\phi$ which is the ratio of the density of $(-)$ particles in lane $1$ and the fixed total density of particles $\rho_{T}\equiv 2\rho_{o}$. We also look at the relative cluster size $\Delta$, defined as the ratio of cluster size in lane $1$ and the length of the lattice $L$.

In the {\it absorbing} phase, for $\mu > \gamma$, the system phase segregates with all the $(+)$ particles occupying lane $1$, while all the $(-)$ particles occupy lane $2$. Correspondingly for $\gamma > \mu$, all $(+)$ particles occupy lane 2 while the $(-)$ particles are all in lane $1$. The density and the current profile are homogeneous and it matches with the Mean Field (MF) results as would be expected for a totally asymmetric exclusion process (TASEP), with  $p_{1} = 2\rho_{o}$, $p_{2} = 0 $,  $n_{1} = 0$, and  $n_{2} = 2\rho_{o}$, while the corresponding currents are, $J_{1}^{+} = 2\rho_{o}( 1 - 2\rho_{o})$, $J_{2}^{+} = 0 $,  $J_{1}^{-} = 0$, and  $J_{2}^{-} = -2\rho_{o}( 1 - 2\rho_{o})$ \cite{ignajstat}. This steady state is an {\it absorbing} state  because once the system gets into this configuration, there is no particle exchange between the lanes, and no microscopic site dynamics can take it out of this state. Further in this phase, $\phi = 0$ while the relative cluster size is of $O(0)$.

\begin{figure}[h]
\centering
\includegraphics[width=2.6in,height = 2.2in, angle=0]{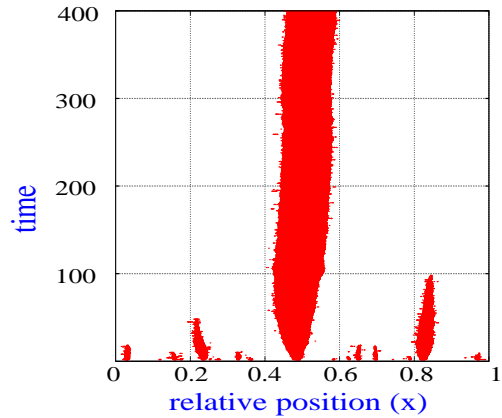}
\begin{center}
\caption{Time evolution of the system leading to formation of the jammed cluster in lane 1, starting from a random initial configuration. Here, $ \rho_{o} = 0.1$, $\beta = 0.15$, $\gamma = 0.4$, $K = 1.2$. Time is in the units of $10^{3}$ MC steps. MC simulation is done for N = 10000.} 
\label{fig-jam}
\end{center}
\end{figure}
\begin{figure}[h]
\centering
\includegraphics[width=2.2in,height = 2.6in, angle=-90]{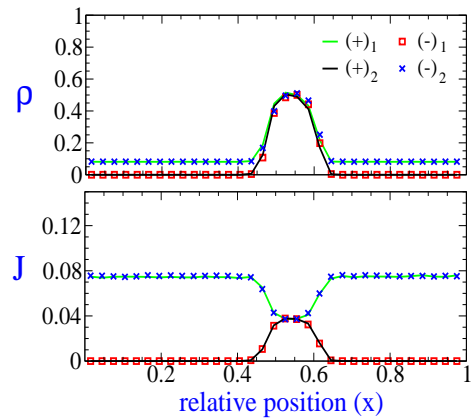}
\begin{center}
\caption{Steady state density$(\rho)$ and current ($J$) profile for $(+)$ and $(-)$ species in lane $1$ and $2$ as function of normalized distance $(x)$ in the jammed phase (corresponding to Fig.1). Parameter values used for MC simulations are the same as in Fig.1.} 
\label{fig-density}
\end{center}
\end{figure}

\begin{figure}[h]
\centering
\includegraphics[width=2.6in,height = 2.6in, angle=-90]{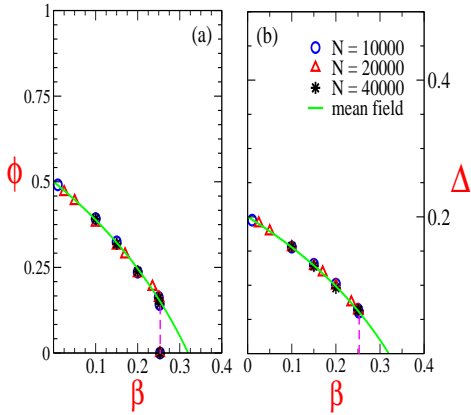}
\begin{center}
\caption{(a)  Variation of $\phi$ with $\beta$: $\phi$ changes discontinuously at $\beta_c = 0.253$. As $\beta \rightarrow 0$, $\phi \rightarrow 0.5$.(b) The relative cluster size $\Delta$, also changes discontinuously at $\beta_c$ and it saturates to a value of $2\rho_{o}$, as $\beta \rightarrow 0$. The MC simulations for different system sizes $N$, overlap with each other and with the MF prediction (solid lines), in the jammed phase. MC simulation are done for parameter values of $K = 1.2$, $\gamma = 0.4$, $\rho_{o} = 0.1$.}
\label{fig-op}
\end{center}
\end{figure}

In the {\it jammed} phase (Fig.1), we observe formation of a single cluster in each of the lanes. This cluster  comprises of both $(+)$ and $(-)$ with no vacancies, with a density of $1/2$ for both the species (Fig.2). In the rest of the region outside the cluster, there is a homogeneous distribution of $(+)$ and vacancies with  absence of $(-)$ in lane $1$, and homogeneous distribution of $(-)$ and vacancies with  absence of $(+)$ in lane $2$. However this apparent phase separation and formation of a {\it condensate}, that we observe in numerical MC simulations (done for system sizes up to $10^{5}$) may not hold true in general in the thermodynamic limit of $N\rightarrow \infty$. This will indeed be the case if distribution of the cluster sizes is such that the mean cluster size is of the same order or large than the system sizes accessed by simulations\cite{schutzrev,kafri1}. In fact for some systems, such apparent {\it condensation} phenomenon was observed in numerical simulations of finite size systems \cite{arndt,korniss}, while subsequently it was shown that phase separation did not exist in the thermodynamic limit of $N\rightarrow\infty$ \cite{kafri1}. Based upon correspondence between the zero range process(ZRP) and driven diffusive models, a numerical criterion was proposed to predict the existence of phase separation in the thermodynamic limit \cite{kafri1}. We differ the discussion about the applicability of this criterion to the model discussed here towards the end. Instead we focus on the jammed steady state of the finite size systems that we can access through simulations. For the single clusters in jammed phase, the MC simulation matches well with the MF value of current; inside the cluster $J_{1}^{+} = J_{2}^{+} = \frac{\beta}{4}$, while outside the cluster  $J_{1}^{+}  = \frac{\beta}{2}$ and $J_{2}^{+} = 0$, with the system being in maximal current phase. The currents for $(-)$ are exactly the same in magnitude with opposite sign inside the cluster, while outside the cluster  $J_{2}^{-}  = \frac{-\beta}{2}$ and $J_{1}^{-} = 0$. The overall total current of $(+)$ and $(-)$, obtained by adding the current in lane 1 and  2, remains constant both inside and outside the cluster (Fig.2). The clusters in both the lanes are co-localized. The densities outside the cluster obtained from MC simulation matches well with the MF expression for density $ p_{1} = 1/2(1 - \sqrt{1 - 2\beta})$, which can be obtained by applying current continuity condition inside and outside the cluster in each lane. The MF expression for the relative cluster size $\Delta$ is obtained by equating the total conserved density of the particles to the individual expression of densities inside and outside the cluster.
\begin{equation}
\Delta = \frac{4\rho_{o} - 1 +  \sqrt{1 - 2\beta}}{1 +  \sqrt{1 - 2\beta}}
\end{equation}

\begin{figure}[h]
\centering
\includegraphics[width=2.6in,height = 2.6in, angle=-90]{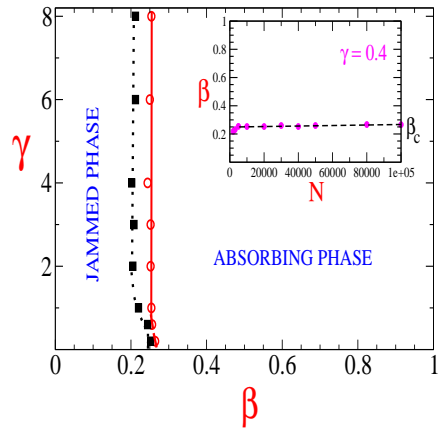}
\begin{center}
\caption{Phase diagram in $\beta-\gamma$ plane for fixed value of $K = 1.2$, $\rho_{o} = 0.1$. The solid lines separates the jammed phase from the absorbing phase.The circles are the data points obtained for starting initial condition of equal density in both lanes. The squares are data points that separates the two phases for an initial condition for which the density of $(-)$ in lane $1$ is $20\%$ of the fixed density $\rho_{o} =0.1$. The inset figure shows the variation of $\beta_c$ with system size for $\gamma = 0.4$, with starting initial condition of equal density of particles in both lanes. MC simulations are done for $N = 50000$.}
\label{fig-phase}
\end{center}
\end{figure}

\begin{figure}[h]
\centering
\includegraphics[width=2.4in,height = 2.4in, angle=-90]{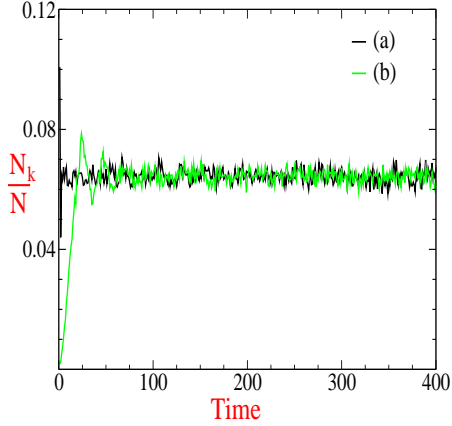}
\begin{center}
\caption{(a) Time evolution of the relative number of kinks $(N_{k}/N)$ for two different initial conditions, (a) equal density of particles in both lanes. (b) 10 $(-)$ particles in lane $1$ and 10 $(+)$ particles in lane $2$. Here, $ \rho_{o} = 0.1$, $\beta = 0.15$, $\gamma = 0.4$, $K = 1.2$. These chosen parameter values corresponds to a point in jammed phase of Fig. 4. MC simulation done for N = 10000. Time is in the units of $10^{3}$ MC steps.}
\label{fig-kink}
\end{center}
\end{figure}
This matches well with the MC simulation results (Fig.3). $\Delta$ takes a value of $2\rho_{o}$ for $\beta\rightarrow 0$. This indicates that in the limit, where $\alpha$ is much faster than $\beta$, all the particles in the lattice tend to accumulate in one large cluster in both the lanes while rest of the lane is vacant. The MF expression for the order parameter is $\phi = \frac{\Delta}{4\rho_{o}}$ and in the limit of $\beta\rightarrow 0$ it assumes a value of $1/2$ indicating that both species of particles are within the cluster and are equally distributed between the two clusters as confirmed by MC simulations.

The entire phase diagram can be specified in terms of lattice hopping rate $\beta$, the lane switching rate $\gamma$, the switching rate constant $K\equiv \frac{\mu}{\gamma}$ and the fixed density $\rho_{o}$. We obtain the phase diagram using MC simulations. In Fig.~4 we show the phase diagram in $\beta- \gamma$ plane for a fixed value of $K$ and $\rho_{o}$. The phase boundary line separating the jammed phase with the absorbing phase depends on the initial starting configuration. The phase boundary for an initial starting condition of equal density of $(+)$ and $(-)$ in both lanes is shifted when compared to an initial condition where initial density of $(-)$ in lane $1$ is $20\%$ of the fixed density $\rho_{o}$ (Fig.~4). Thus the phase boundary appears as a narrow band of region in the phase plane rather than a sharp line, indicating that self-averaging does not happen in the vicinity of the transition boundary. This might be an artifact arising out of finite size effects. However for the condition of same specified initial density in each lane, phase transition point $\beta_{c}$ remains unchanged for different system sizes (Fig.~4 inset). We note that deep inside a particular phase (beyond the region of narrow band), the final steady state is independent of the initial configuration and it is uniquely determined in terms of the density, current, and the order parameter $\phi$ corresponding to that particular phase. To illustrate this point, we define kink number $N_k$ which corresponds to the total number of times a $(+)$ is followed by a $(-)$  in the same lane. Fig.~5 shows the temporal evolution of the relative kink number ($N_k/N$) for very different initial conditions, where the final steady state corresponds to the same jammed phase. In fact for the {\it absorbing state}, $N_k$ is zero  while $N_k/N$ is a finite value, whose average value is independent of the system size in the jammed phase. Further, we have checked that the relative fluctuations of the relative kink number decreases with the system size, indicating that the system gets kinetically trapped in the jammed state in the thermodynamic limit of $N\rightarrow \infty$. 
In order to understand what determines the phase boundary between the jammed and the absorbing phase, we look at the temporal behavior of the system in the vicinity of the absorbing state. In particular, we perform a linear stability analysis of the MF steady state fixed point corresponding to the absorbing steady state. The MF evolution equations for the system can be expressed in terms of the mean site occupation densities \cite{ignajstat},
\begin{eqnarray}
\partial_{t}p_{1}& =&\mu p_{2}n_{2}( 1 - p_{1} - n_{1}) -  \gamma p_{1}n_{1}( 1 - p_{2} - n_{2}) \nonumber\\
&+&\epsilon\left[\mu p_{2}( 1 - p_{1} - n_{1})\partial_{x} n_{2} -\gamma p_{1}( 1 - p_{2} - n_{2})\partial_{x} n_{1} \right]\nonumber\\
&-&\epsilon\partial_{x}\left[\alpha p_{1}( 1 - p_{1} - n_{1}) + \beta p_{1}n_{1}\right] + O(\epsilon^{2})
\label{mf-eqn1}
\end{eqnarray}

\begin{eqnarray}
\partial_{t}p_{2}& =&\gamma p_{1}n_{1}(1 - p_{2} - n_{2}) - \mu p_{2}n_{2}( 1 - p_{1} - n_{1}) \nonumber\\
&+&\epsilon\left[\gamma p_{1}( 1 - p_{2} - n_{2})\partial_{x} n_{1} - \mu p_{2}( 1 - p_{1} - n_{1})\partial_{x} n_{2}\right]\nonumber\\
&-&\epsilon\partial_{x}\left[\alpha p_{2}( 1 - p_{2} - n_{2}) + \beta p_{2}n_{2}\right] + O(\epsilon^{2})
\label{mf-eqn2}
\end{eqnarray}

\begin{eqnarray}
\partial_{t}n_{1}& =&\gamma p_{2}n_{2}( 1 - p_{1} - n_{1}) -  \mu p_{1}n_{1}( 1 - p_{2} - n_{2}) \nonumber\\
&+&\epsilon\left[\mu n_{1}( 1 - p_{2} - n_{2})\partial_{x} p_{1} -\gamma n_{2}( 1 - p_{1} - n_{1})\partial_{x} p_{2} \right]\nonumber\\
&+&\epsilon\partial_{x}\left[\alpha n_{1}( 1 - p_{1} - n_{1}) + \beta p_{1}n_{1}\right] + O(\epsilon^{2})
\label{mf-eqn3}
\end{eqnarray}

\begin{eqnarray}
\partial_{t}n_{2}& =&\mu p_{1}n_{1}( 1 - p_{2} - n_{2}) - \gamma p_{2}n_{2}( 1 - p_{1} - n_{1}) \nonumber\\
&+&\epsilon\left[\gamma n_{2}( 1 - p_{1} - n_{1})\partial_{x} p_{2} - \mu n_{1}( 1 - p_{2} - n_{2})\partial_{x} p_{1}  \right]\nonumber\\
&+&\epsilon\partial_{x}\left[\alpha n_{2}( 1 - p_{2} - n_{2}) + \beta p_{2}n_{2}\right] + O(\epsilon^{2})
\label{mf-eqn4}
\end{eqnarray}
where we have displayed terms up to first power of $\epsilon$.

For $\mu > \gamma$, the homogeneous MF steady state solution for the density is $p_{1} = 2 \rho_{o}, p_{2} = 0, n_{1} = 0$ and $n_{2} = 2 \rho_{o}$.
\begin{figure}[h]
\centering
\includegraphics[width=2.4 in,height = 2.6in, angle=-90]{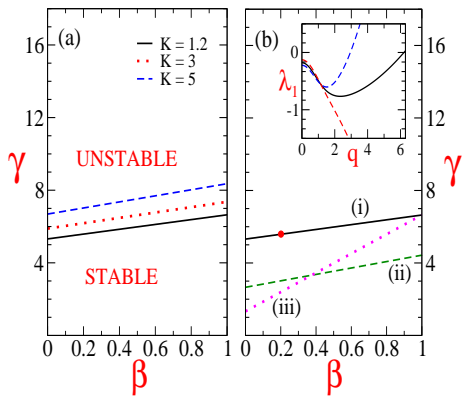}
\begin{center}
\caption{(a) MF linear stability line as function of $K$: Increasing $K$ tends to increase the region of homogeneous {\it absorbing} phase. Here $\rho_{o} = 0.1$. (b) Effect of variation of $\rho_{o}$: (i) $\rho_{o} = 0.1$,(ii) $\rho_{o} = 0.2$, (iii) $\rho_{o} = 0.4$. Here $K= 1.2$. Inset shows the marginally stable mode (solid line) corresponding to the circle on the stability line in Fig. 6(b), and the effect on mode stability by variation of $\gamma$. Here $\beta = 0.2$.}
\label{fig-mf}
\end{center}
\end{figure}

For performing linear stability analysis about the homogeneous MF steady state, we have to take into account the terms in Eqn.(\ref{mf-eqn1}-\ref{mf-eqn4}) which are $O(\epsilon^{2})$. Following the usual procedure of retaining terms up to linear order in fluctuations of the variables, $p_1$, $p_2$, $n_1$ and $n_2$ and taking spatial Fourier transforms of the fluctuations; $\delta p_1$, $\delta p_2$, $\delta n_1$ and $\delta n_2$, we obtain the corresponding eigenvalues, which determines the stability of the MF homogeneous phase. The corresponding eigenvalues are,
 \begin{eqnarray}
 \lambda_{1,2}& = &-i\epsilon qM - \left[ \mu A + \frac{1}{2}\epsilon^{2}q^{2}M \mp \gamma A{\left(1 + \frac{1}{4}\epsilon^{4}q^{4}\right)}^{\frac{1}{2}}\right] \nonumber\\
 \lambda_{3,4}& = &\pm i\epsilon q\alpha(1 - 2\rho_{o}) - \frac{1}{2}\epsilon^{2}q^{2}\alpha 
 \label{mf:eigenvalue}
 \end{eqnarray}
 Here, $M = \alpha(1 -2\rho_{o}) + 2\rho_{o}\beta$ and $A = 2\rho_{o}(1 -2\rho_{o})$.

Of the four eigenvalues, only for $\lambda_{1}$ the real part of the eigenvalue can be positive depending on the values of the parameters and wave number $q$. The other three eigenvalues always correspond to stable modes of fluctuations. Fig.~6(inset) shows the variation for $\lambda_{1}$ as a function of wave number $q$. In fact the long wavelength limit ($q = 0$) fluctuation is always stable as $\lambda_{1}(q =0) = (-\mu + \gamma)A$ is always negative for $\mu > \gamma$. For certain critical value of $q$, $\lambda_{1}$ becomes positive. However the maximum value of wave number $q_m$, that is possible is limited by the lattice spacing, so that $q_{m} = \frac{2\pi N}{L}$, corresponding to a fluctuation at the scale of one lattice spacing. Now the expression for the stability line can be  obtained by substituting the expression $q_{m}$ in Eqn.(\ref{mf:eigenvalue}) and equating it to zero. Setting $\alpha$ to $1$, we obtain the equation for MF linear stability line,
\begin{equation}
\gamma = 2\pi^{2}\frac{\frac{1}{2\rho_o} + \frac{\beta}{1 - 2\rho_o}} {{(1 + 4\pi^{4})}^{1/2} - K }
\end{equation}

Fig.~6(a) and Fig.~6(b) shows the variation of the position of the MF stability boundary with $K$ and $\rho_{o}$ respectively. Comparing the MF stability line of Fig.~6(a) (solid lines), with the numerical phase boundary in Fig.~4, we can see that it does not agree with numerical simulation result. Since the MF stability line is determined by the instabilities of large wavenumber fluctuations, it is only expected that at the scale of lattice spacing, the correlations of fluctuations between the neighbouring lattice sites cannot be neglected, leading to inaccuracies in MF analysis. So while the MF analysis predicts the instability of the homogeneous steady state for certain range of parameters, it fails to capture the location of the phase boundary.       

Finally we look at the issue of formation of {condensate} in the jammed phase in the thermodynamic limit. Many models which carry a non-vanishing current in the thermodynamic limit, the current in a finite cluster of size $n$ takes an asymptotic form $J_{n} \sim J_{\infty}(1 + b/n^{\sigma})$ to leading order in $1/n$ \cite{kafri1,kafri2}. Using a correspondence between asymptotic form of current in the cluster for zero-range process (ZRP) and such models, it has been proposed that phase separation leading to formation of a single condensate can occur for either $\sigma<1$ and $b>0$ or for $\sigma=1$ and $b>2$ \cite{kafri1}. This conjecture has been applied for a two lane model \cite{korniss}, by performing MC simulation for the open two lane system of size $n$, without vacancies and with {\it equal} rate of particle entry and exit at the boundaries. It has been used to determine the finite size corrections to current $\Delta_{n}= (J_{n} -J_{\infty})/J_{\infty}$ and extract the corresponding values of $\sigma$ and $b$ \cite{kafri1}. However there are two issues that we wish to highlight when we apply this criterion for our case: (i) The region adjoining the cluster comprises of fluid phase of $(+)$ alone, while for lane $2$ this region is solely a fluid phase of $(-)$ (Fig.2). Thus it is {\it a priori} not clear whether the simulation for the open system should be performed with equal entry and exit rates of the particles at the boundaries. We perform the MC simulations for both cases, e.g; with {\it equal} boundary rates of entry and exit for each particles in both the lanes, and with no entry of $(-)$ particles in lane $1$ and no entry of $(+)$ particles in lane $2$; (ii) In the MC simulations, we find that the root mean square(RMS) fluctuations of the measured current $\delta J_{n} > \Delta_{n}$ \footnote{For $N=10^3$, $\delta J_{n}\simeq 0.005$ while $\Delta_{n}\simeq 0.0024$ and for $N= 10^4$, $\delta J_{n}\simeq 0.0015$ while $\Delta_{n}\simeq 0.0003$.}. Further we find that increasing the iterations for obtaining the average current does not significantly change the RMS fluctuation of $J_n$. This implies that the estimates of $b$ and $\sigma$ obtained from fitting the data would be rather unreliable. In Fig.~7 we show the straight line fit ( with $ \sigma =1$) for the data points obtained for unequal entry rates of particles in two lanes corresponds to $b=2.86$. For the data points corresponding to equal entry and exit rates, $b=3.03$. Thus both of these data set  suggests {\it condensation}. If we fit with $\sigma\neq1$, we obtain $\sigma<1$ for both data sets, suggesting again the same conclusion. However owing to the limitations of high $\delta J_{n}$ which is larger than $\Delta_{n}$, we cannot definitively conclude the existence of phase separation and formation of single condensate for $N\rightarrow\infty$ based on these simulation results.
\begin{figure}[h]
\centering
\includegraphics[width=2.3 in,height = 2.3in, angle=-90]{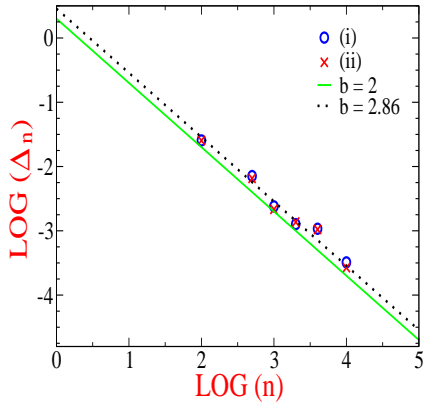}
\begin{center}
\caption{Finite size corrections to current $\Delta_{n}$:(i) Equal entry and exit rate of particles for both lanes ($\xi$).(ii) No input of $(-)$ particles in lane $1$ and $(+)$ particles in lane $2$, the other input and exit rates being ($\xi$).  Solid line corresponds to $b=2$ with $\sigma=1$, while the dotted line is the best fit (with b = 2.86), obtained with data point of (ii) with $\sigma =1$. Here $K = 1.2$, $\alpha = 1$ and $\xi =1$.}
\label{fig-finitesize}
\end{center}
\end{figure}   

In summary, we have studied an unique jamming transition between a single species homogeneous absorbing state and a clustered jammed phase comprising of both the species in driven lattice gas system. Simulations based on the criterion proposed in Ref.\cite{kafri1} does not conclusively resolve whether the jammed state is single {\it condensate} in the thermodynamic limit. This is due to the limitations placed by the relatively high values of the fluctuations of current when compared with the finite size correction to current $\Delta_n$. The MF theory within a particular phase is able to accurately predict the steady state profile, although the transition itself is not well described by a MF analysis.

While in this letter we have focused on discussing the statistical mechanics aspect of the model, this minimal model may provide insight to jamming phenomenon arising out of the interplay of translation processes of motors on cellular filaments and their lane switching dynamics. This may be relevant in understanding jamming phenomenon that is seen in context of axon transport and manifests in the form of neurodegenerative diseases like Alzheimer's.

\section{Acknowledgment}
I would like to thank Deepak Dhar for useful discussions and suggestions.

\end{document}